\documentclass[aps,pre,showpacs,twocolumn,superscriptaddress,floatfix]{revtex4}
\usepackage{graphicx}
\usepackage{epsfig} 
\usepackage{epstopdf}
\usepackage{color}
\usepackage{amsfonts}
\usepackage{amssymb,amsmath}
\usepackage{bbold}

\begin{document}
 
\title{Complete synchronization equivalence in asynchronous and delayed coupled maps}

\author{Juan Carlos Gonz\'alez-Avella}
 \email{avellaj@gmail.com}

\author{Celia Anteneodo}
 \email{celia.fis@puc-rio.br}

\affiliation{Department of Physics, PUC-Rio, Caixa Postal 38071,
  22452-970, Rio de Janeiro, Brazil} \date{\today}

\begin{abstract}

Coupled map lattices are paradigmatic models of many collective phenomena. 
However, quite different patterns can emerge depending on the updating scheme. 
While in  early versions, maps  were updated synchronously, 
there is, in recent years, a concern to consider more realistic updating schemes, 
where elements do not change all at once. 
Asynchronous updating schemes and the inclusion of time delays 
are seen as more realistic than the traditional parallel dynamics, 
and, in diverse works, either one or the other, have been implemented separately. 
But are they actually distinct cases?   
For coupled map lattices with adjustable range of interactions, 
we prove, using both numerical and analytical tools,  that an adequate delayed 
dynamics leads to the same completely synchronized states that an asynchronous update, 
providing a unified framework to identify the stability conditions for complete synchronization.

 \end{abstract}

\maketitle

\section{Introduction}

An emblematic example of the spatiotemporal collective properties that can 
arise in  ``complex systems'' is provided by the phenomenon of synchronization. 
Such emergent behavior can be found in many real world systems, e.g.,  laser arrays,  
pacemaker heart cells, circadian rhythms, flashing  fireflies, amongst many 
others~\cite{PykovskyBook,ManrubiaBook,KuramotoBook}. Simple models that capture 
the essential features of the synchronization phenomenon~\cite{KanekoBook}   
are coupled map lattices (CMLs). Moreover, they are useful to model systems 
of scientific and technological interest as diverse as  Josephson junction arrays, 
multimode lasers, vortex dynamics, even evolutionary biology~\cite{PykovskyBook}, 
and also constitute good prototypes to investigate control of chaos. 

Traditionally, when studying emergent patterns in CMLs,  
the update of the constituent units is made synchronously (parallel updating), 
meanwhile,   the elements in real arrays are not perfectly synchronous. 
This issue is particularly important, for example, in biological neural 
networks~\cite{HertzBook}, but, more generally, when modeling real systems in diverse contexts, 
it seems more adequate to take into account some kind of asynchronicity. 
This choice is crucial, with important consequences on the final collective 
states~\cite{Sinha2007,Sinha_2000,Lumer1994,nuno1,nuno2}.  For instance, 
asynchronous updating may open windows in parameter space where synchronization 
becomes allowed~\cite{Sinha_2000}  and induce regularity in coupled 
systems in contrast to a synchronous updating~\cite{Sinha2007,unstableFP1,squeeze}.  
Alternatively,  the introduction of  time delays,  
 to account for  finite delays in information transmission between 
units~\cite{Marti_PRL05,Marti_EPJB05,Marti_PRE05,delayed},    
has also a noticeable impact on  the collective patterns, e.g.,  
although synchronization is still possible, chaos is suppressed~\cite{Marti_PRL05}.

Another  realistic ingredient in modeling extended systems, 
within the spatial domain, is the coupling range.  
As a matter of fact, the range of the interactions plays a 
crucial role in the determination of the emergent patterns in any extended 
system~\cite{chemical1,Anteneodo2004,Anteneodo2006,Marti_EPJB05,gallas2004,celia1998,celia2000,lind04}.  
Insofar as the range of the interactions can affect the  propagation 
of information, it is important to explore its interplay with the updating scheme.  

Asynchronous updating, as well as time-delayed interactions, seem natural choices 
for a realistic modeling of coupled  systems,  but the correspondence  
between both descriptions has not been investigated yet. In this work, 
we demonstrate, through  numerical and analytical tools, that the complete synchronization of a CML 
with asynchronous updating can be emulated by a delayed dynamics. 
Additionally, through the range of the interactions, we will show the interplay of the temporal 
and spatial dimensions. 
We will derive analytical conditions for complete synchronization valid for a large class of CMLs and local dynamics.

\section{The Model} 
\label{sec:model}

We consider an array of $N$ elements with periodic boundary conditions.  
The coupled elements evolve according to the mapping 
 \begin{equation}
   x^i  \mapsto (1-\epsilon) f(x^i  ) +  \epsilon 
   \sum_{r=1}^{N'} A(r) \Bigl(  f(x^{i-r} ) +f(x^{i+r} ) \Bigr), 
   \label{eq:CML}
 \end{equation}
for  $i=1,\ldots,N$, where $x^i$ describes the state of element $i$, whose local 
dynamics is governed by the chaotic map $x \to f(x)$. 
This is a fully connected array where elements interact, through a coupling also given by $f(x)$, 
with intensity $A(r)$, where $r$ is the integer inter-element  distance over the ring. 
Parameter $\epsilon$ ($0\le \epsilon \le 1$)  governs the balance between global and local influences.

In numerical examples, we will consider $A(r)= r^{-\alpha}/\eta$,   
 where $\alpha$ $\in [0,\infty)$ determines the range of the interactions, 
and $\eta(\alpha)  = 2\sum_{r=1}^{N'}r^{-\alpha}$ is a normalization factor, 
	where $N' = (N-1)/2$ for odd $N$.
This coupling scheme allows to scan continuously from  global  ($\alpha = 0$) 
to  nearest-neighbor ($\alpha \rightarrow \infty$) interactions. 
Additionally, in numerical simulations, we will use, as paradigmatic example,  the  logistic map	$x \to f(x)=4x(1-x)$.  
However, the analytical expressions that we will derive are valid for generic $A(r)$ and  chaotic unimodal maps $x \to f(x)$.

In the usual synchronous evolution, all the $N$ new states (at discrete time $t$) 
are computed in parallel from the $N$ previous values (at time $t-1$), 
that is, all the sites are updated simultaneously. 
Alternatively, we will also investigate the asynchronous (i.e., random sequential) evolution, 
where the updates are not simultaneous. 
Finally, we will consider also  time-delayed evolutions of the array, 
by introducing random or fixed delays in Eq.~(\ref{eq:CML}).

\section{Results} 
\label{sec:results}

We performed numerical simulations of the CML defined by Eq.~(\ref{eq:CML}), using the different evolution protocols, 
starting from random initial conditions. 
The patterns that emerge through each updating  will be compared, as a function of the 
coupling strength $\epsilon$ and the range of the interactions ruled by $\alpha$.

We monitor the collective behavior by means of the
instantaneous mean field $h$ defined as~\cite{Marti_PRL05,lind04} 
 \begin{equation}
    h_t = \frac{1}{N}\sum_{i=1}^N  x^i_t  \,,
\end{equation}
and use the  time average, $ \langle \sigma \rangle$, of its instantaneous standard deviation  
\begin{equation}
      \sigma_t = \sqrt{ \frac{1}{N}\sum_{i=1}^N 
			\bigl(    x^i_t   -h_t \bigr)^2 } \,,
 \end{equation}
to measure the degree of  synchronization.   
When  $\langle \sigma \rangle = 0$, it means that the system is  
completely synchronized (CS), i.e,   $x_1=x_2=\ldots x_N=x^\star$, where 
$x^\star$ can evolve in a chaotic or in a regular trajectory. 
Another relevant parameter, that allows to characterize the chaoticity of a dynamical system,  
is the largest Lyapunov exponent $\lambda_{\rm max}$~\cite{eckman-ruelle}. 
If $\lambda_{\rm max}$ is  positive, the system displays a chaotic behavior, 
while if it is  negative, the dynamics is in a regular regime.  
We compute $\lambda_{\rm max}$ using the Benettin algorithm~\cite{Benettin_01}.

We will show below the results for each updating scheme.

\subsection{ Synchronous updating}
\label{sec:syn}
%
 We first review, as a reference that will be useful later, the well known case where all the maps are updated at once, 
in which case     
Eq.~(\ref{eq:CML}) reads 
 \begin{equation}
   x^i_t  = (1-\epsilon) f(x^i_{t-1}) +  \epsilon 
   \sum_{r=1}^{N'} A(r) \Bigl(  f(x^{i-r}_{t-1} ) + f(x^{i+r}_{t-1} ) \Bigr). 
   \label{CMLs}
 \end{equation}
In a CS state, the mapping (\ref{CMLs}) becomes $ x^\star_t  = f(x^\star_{t-1})$, for all $i$, 
that is, the nonlocal influences vanish, and each map evolves with the uncoupled  local chaotic dynamics. 
However, the array parameters $\epsilon$ and $\alpha$ participate in determining the stability of the CS state.

The domain of CS can be obtained analytically as follows (see for instance Refs.\cite{Anteneodo2004,lind04}). 
Linearizing Eq.~(\ref{CMLs}) around the CS state 
(where all maps are in a state $x^\star_t$ at time $t$), 
one gets the map of small displacements 
$\mathbf{\delta x}_{t+1} = \mathbf{F}_t\mathbf{\delta x}_{t}  $, 
where $\mathbf{F}_t$ is the $N\times N$ matrix 
\begin{equation}
\mathbf{F}_t = [ (1-\epsilon) \mathbb{1} + \epsilon \mathbf{A} ]f'(x^\star_t)  \,,
\label{matrixF}
\end{equation}
with    $A_{ij}=(1-\delta_{ij}) A( r_{ij})$, being $r_{ij}={\rm min}_k|i-j+kN|$. 
Moreover, recall that  the Lyapunov exponent of the uncoupled map is 
${\rm e}^{\lambda_u}= \lim_{t \to \infty} \prod_{n=0}^{t-1}|f'(x^\star_n)|^{1/t}$, 
computed over a chaotic trajectory. 
Therefore,  the chaotic synchronized state is stable, if the eigenvalues of the matrix
${\rm e}^{\lambda_u}[ (1-\epsilon) \mathbb{1} + \epsilon \mathbf{A} ] $, related to transverse eigenvectors,  
are smaller than one in absolute value. This means
\begin{equation}
-1\le {\rm e}^{\lambda_u}[1-\epsilon(1-a_k)]  \le 1, \;\;\;\mbox{for all $k<N$}, 
\label{frontiers}
\end{equation}
where $a_k=\sum_{m=1}^{N'} A(m) \cos(2\pi km/N) $, 
for $1\le k \le N$ are the eigenvalues of $\mathbf{A}$, 
that can be obtained by Fourier diagonalization~\cite{Anteneodo2004}.
The  domain of chaotic synchronization defined by the double inequality (\ref{frontiers}) 
is the region in the plane $\epsilon-\alpha$ below the dashed line  in  Fig.~\ref{fig:PDa}.a. 
This region agrees perfectly with the region where $\langle \sigma \rangle =0$ in numerical simulations (not shown). 
In this domain $\lambda_{\rm max}$ takes the positive value of the chaotic uncoupled map 
(i.e., $\lambda_{\rm max}= \lambda_u = \ln 2$ in our case), because the maps in the array follow the local dynamics. 
The critical strength $\epsilon_c$ increases with $\alpha$, hence, the 
synchronization interval shrinks and collapses  at $\alpha\simeq 0.8$. 
Therefore,  too short-range interactions are not able to fully synchronize the system~\cite{Sinha_2000,Anteneodo2004}.

\subsection{Asynchronous updating}
\label{sec:asyn}

 In this case,  we proceed as follows:

(i) An element of the array is selected at random.

(ii) Its state is updated according to Eq.~(\ref{eq:CML}), using 
the most recent state  of the array.

(iii) After $N$ iterations (updates), the discrete time is increased in one unit.

\begin{center}
\begin{figure}[h!]
\includegraphics[width=0.9\linewidth,angle=0]{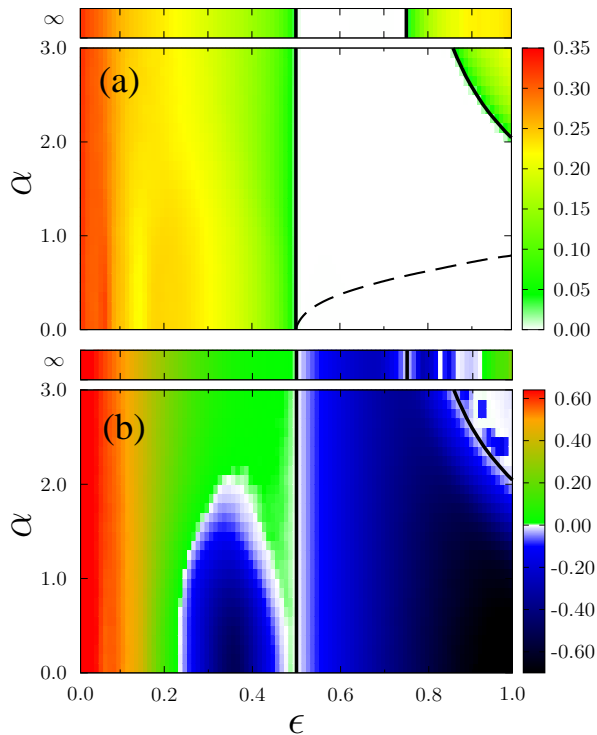}
\caption{ (Color online) {\bf Asynchronous updating}.
Phase diagrams, 	in  the parameter plane $\epsilon-\alpha$, showing 
	$\langle \sigma \rangle$ (a) and   $\lambda_{\rm max}$ (b)  in color scale.  
The black solid lines correspond to the theoretical prediction given by Eq.(\ref{frontiers2}). 
The dashed line, that delimits the region of chaotic CS in the synchronous case {\it A}, is given by 
~Eq.~(\ref{frontiers}).
In all cases, the array size is  $N=1000$.
For each value of $\epsilon$,  we used 100 time steps, 	after $ t=10^4$ has elapsed, over a typical trajectory, 
starting from random initial configurations where each $x^i$ is random in [0,1],   
to compute $\langle \sigma \rangle$ and $\lambda_{\rm max}$. 
}
\label{fig:PDa}
\end{figure}
\end{center}

We performed numerical simulations of the asynchronous dynamics defined above, for arrays of size $N$
(essentially the same results are observed for sizes $N\gtrsim 200$) and different values of the parameters.  
The effect of the interaction range parameter $\alpha$ is depicted continuously in the phase diagrams that show     
$\langle \sigma \rangle$ (Fig.~\ref{fig:PDa}.a) and $\lambda_{\rm max}$ (Fig.~\ref{fig:PDa}.b), 
on the  plane ($\epsilon,\alpha$). 
The CS domain (white region)  is enhanced 
with respect to the synchronous case (below dashed curve), for $\alpha>0$.  
When $\alpha$ increases, a window of CS,  $\epsilon_c < \epsilon <\epsilon_c^\prime$, 
persists  for any $\alpha$    in the asynchronous case. 
While  $\epsilon_c$ remains constant,  $\epsilon_c^\prime$ diminishes with $\alpha$,  above $\alpha=2$, 
but a window of CS survives even in the limit $\alpha\to\infty$,   
in accord with results previously reported for nearest-neighbor interactions~\cite{Sinha_2000}.

\begin{center}
\begin{figure}[h!]
\includegraphics[width=0.9\linewidth,angle=0]{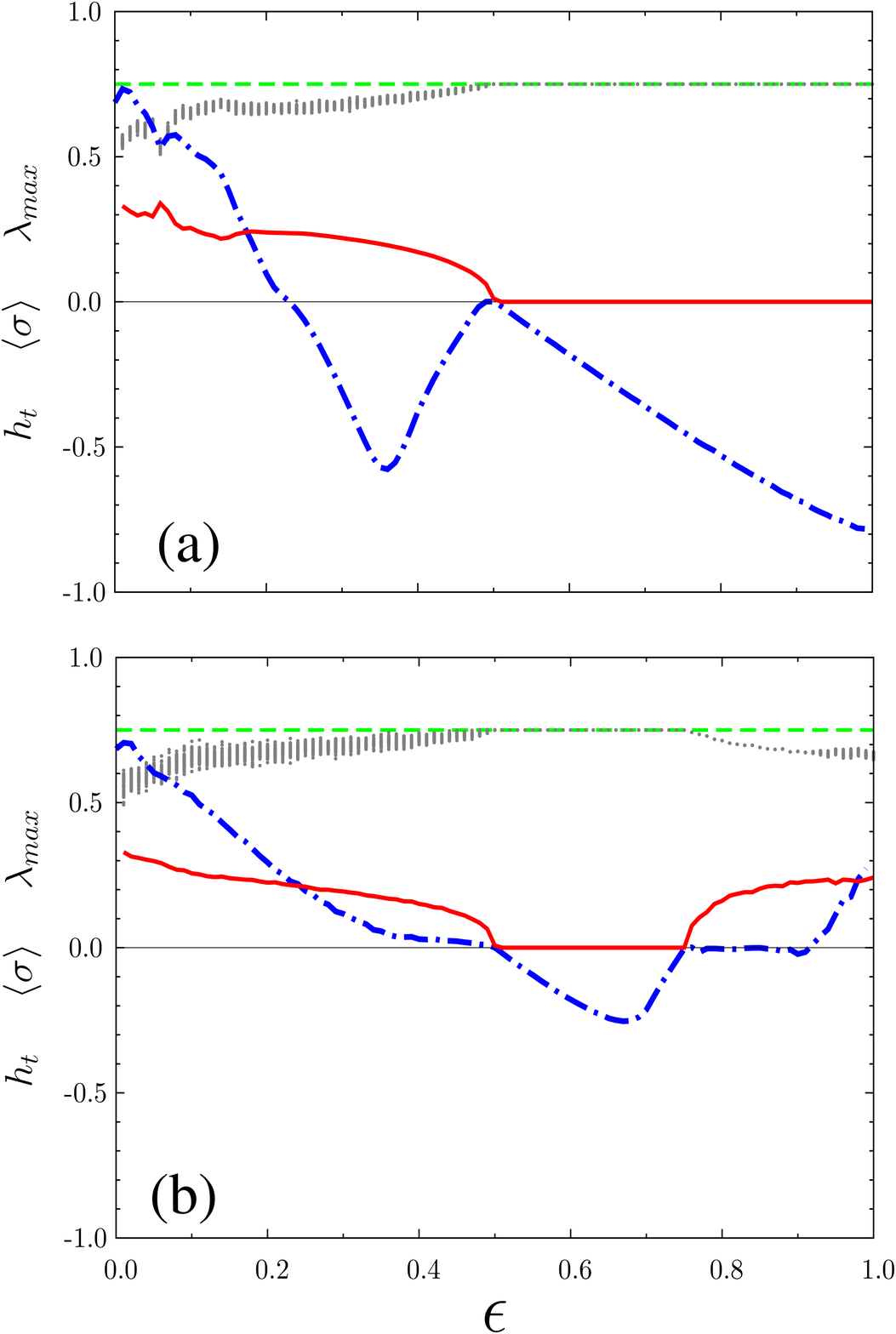}
\caption{ (Color online) {\bf Asynchronous updating}.
Cuts of the phase diagrams in Fig.~\ref{fig:PDa}, at $\alpha = \infty$ (a) and $\alpha = 0$ (b), 
showing $\langle \sigma \rangle$ (red solid line), $\lambda_{\rm max}$ (blue dash-dot line) and, 
additionally, the order parameter $h_t$ (light-gray dots),    all vs. the coupling $\epsilon$. 
The green dashed line, in   (a) and (b), corresponds to the  fixed point at $h_t=3/4$, plotted as a reference.  
}
\label{fig:hta}
\end{figure}
\end{center}

The bifurcation diagrams of $h_t$, as well as  $\langle \sigma \rangle$ and $\lambda_{\rm max}$,  
vs the coupling strength $\epsilon$ are shown in Fig.~\ref{fig:hta},  
in the extreme cases  $\alpha \rightarrow \infty$ (a) and $\alpha = 0$ (b). 
First we notice that even in the case of nearest neighbors ($\alpha \to \infty$), CS can emerge 
(pointed by intervals of $\epsilon$ where  $\langle \sigma \rangle=0$). But, in those regions,  
the negative $\lambda_{\rm max}$ indicates that, differently from the synchronous case, CS is non-chaotic. 
In fact, synchronization occurs at $x^\star=3/4$, that corresponds to a fixed point of the logistic map, 
as evinced by the plot of $h_t$. 
Chaos is suppressed in all the CS domain    
such that the collective state of the system is always the spatially homogeneous one given by $x^\star=3/4$. 
This is the unstable fixed point of the individual map, that gained stability in the coupled system~\cite{squeeze}.

We can derive the critical values of $\epsilon$, under the asynchronous updating,  
by analyzing first the stability of the fixed point under a synchronous dynamics. 
In the latter case, the stability condition of a CS solution at a fixed point 
can be obtained analytically from  the eigenvalues $\phi_k$, $k=1,\ldots,N$ of the matrix $\mathbf{F}$ defined in 
Eq.~(\ref{matrixF}). Differently from case {\it A},    in the fixed point $x^\star=3/4$  where 
$f'(x^\star)=-2$, $\mathbf{F}$ is  time independent. 
For stability, we must request  that $|\phi_k|\le 1$ for $k=1,\ldots, N-1$. 
Therefore the condition for transverse stability is 
the same   given by Eq.~(\ref{frontiers}). 
However, this stability is transversal, while,  in case {\it A}, $x^\star=3/4$ is unstable along the synchronization subspace, 
therefore the dynamics becomes chaotic, as seen in the previous subsection. 
In the asynchronous case, the scenario is rather different, as shown in Fig.~\ref{fig:PDa}, 
and can be heuristically understood as follows~\cite{unstableFP1}. 
Near the lower frontier $\epsilon_c$, where the eigenvalues are close to -1, 
and therefore the individual deviations from the fixed point 
change sign at each iteration, the nonlocal contribution is expected to cancel out. 
Therefore, the stability condition that the eigenvalues must fulfill becomes $-1 \le -2(1-\epsilon)$, 
implying $\epsilon_c=1/2$ for all $\alpha$.   
Differently, in the upper frontier  $\epsilon_c^\prime$,  such cancellation does not take place, then 
the inequality (\ref{frontiers}) for $\epsilon_c^\prime$ in the synchronous case must still hold.    
Therefore, 
\begin{equation}
1/2=\epsilon_c \le \epsilon \le \epsilon_c^\prime= 3/\left( 2(1-{\rm min}[\{a_k\}]) \right),
\label{frontiers2}
\end{equation} 
which is in excellent accord with numerical outcomes,  
as can be seen in  Fig.~\ref{fig:PDa}.a.

 \subsection{Delayed dynamics}
\label{sec:delayed}

Now we consider the following time-delayed CML 
 \begin{equation}
   x_{t+1}^i = (1-\epsilon) f(x_t^i) +  \epsilon 
   \sum_{r=1}^{N'} A(r) \Bigl( f(x_{t-\tau_{i,i-r}}^{i-r}) +
     f(x_{t-\tau_{i,i+r}}^{i+r}) \Bigr),
   \label{CMLtau}
 \end{equation}
where the whole number $\tau_{i,j}$ is the delay of element $i$ in response to element $j$. 
The updating of the set of maps is performed simultaneously.

\begin{center}
\begin{figure}[b!]
\includegraphics[width=0.9\linewidth,angle=0]{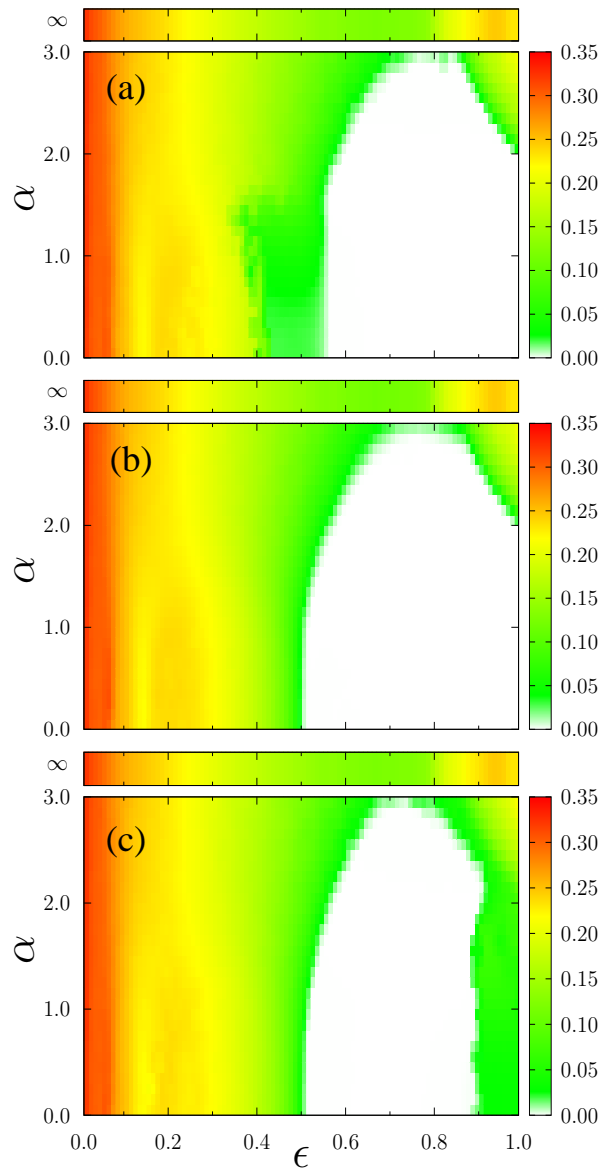}
\caption{(Color online)  {\bf Time-delayed dynamics}. 
	Phase diagrams, 	in  the parameter plane $\epsilon-\alpha$, showing 
	$\langle \sigma \rangle$  in color scale, for the binary distribution ($\tau= 1,0$, with  probabilities $p$, $1-p$), 
	with  $p=0.45$ (a),	 $p=0.5$ (b) and $p=0.55$ (c).  
White regions indicate CS states, where $\langle \sigma \rangle=0$. 
	}
\label{fig:PDp}
\end{figure}
\end{center}

We analyzed different distributions of delays, where $\tau_{ij}$ can be fixed or randomly chosen at each time $t$ 
with a given probability distribution. We report below the cases that we found better match with 
the asynchronous case {\it B}.

First, we consider a binary distribution where $\tau$ can take solely the values 1 and 0, 
with probabilities $p$ and $1-p$, respectively. Hence $p=0$ recovers 
the synchronous case  {\it A}, while $p=1$ corresponds to a dynamics with fixed delay $\tau=1$.

Phase diagrams of $\langle \sigma \rangle$ for  values of $p\approx 0.5$ are 
shown in Fig.~\ref{fig:PDp}. 
For $0\le \alpha \lesssim 1$ the portraits shown in Figs.~\ref{fig:PDa}.a and \ref{fig:PDp}.b ($p=0.5$)
are equivalent, in the sense that the intervals of $\epsilon$ for non-chaotic CS are coincident.  
In fact, the case that better approaches the asynchronous updating scenario B is provided by 
 $p=0.5$ (see Fig.~\ref{fig:PDp}), which yields the largest domain of CS in the plane  $\epsilon-\alpha$. 
Furthermore, like in case {\it B},  the chaotic dynamics  is  suppressed with CS  occurring at $x^\star=3/4$.
However, for $\alpha>1$,  differently to the asynchronous case where $\epsilon_c$  remains constant (Fig.~\ref{fig:PDa}.a), 
now $\epsilon_c$ increases with $\alpha$ (Fig.~\ref{fig:PDp}.b). 
As a consequence,  the CS domain shrinks  and collapses at $\alpha \simeq 3$. 
Then,  for   $\alpha \gtrsim 3$, CS never occurs in the binary delayed case. 
That is, the equivalence of CS domains for  the delayed and asynchronous dynamics fails for short-range couplings.
In fact, while long-range couplings favor  homogenization, through spatial averaging, 
 short-range couplings promote  the formation of  spatial domains of synchronized and  
non-synchronized regions. These spatial patterns are persistent, hindering CS. 
Some sort of temporal average might compensate the lack of spatial homogenization, 
through the interplay between temporal  and spatial dimensions. 
However, binary delays do not perform well that task when short-range interactions are involved.
We also tested other (non binary) distributions of discrete delays, without finding improvements 
in the sense of mimicking the asynchronous dynamics.


If the asynchronous updating can be thought as a sort of  delayed dynamics, 
as far as a distribution of non-integer delays is behind, 
integer values of $\tau_{ij}$ fail  in providing  equivalence of CS 
when the range of the interactions is too short. 
A question is whether that in-equivalence arises due to the discrete nature of time delays. 
Since the states are accessible at discrete times only, 
a way to emulate continuous time delays in the real interval $[0,1]$  
is by interpolation between the states at $\tau=0$ and $\tau=1$, as if the 
trajectory were continuous by parts. Namely, we consider the intermediate state
\begin{equation}
\hat{x}_t= \beta x_{t-1} + (1-\beta) x_{t-0} \,,
\label{beta}
\end{equation}
where  $\beta$ controls the interpolation point and can take  fixed or distributed values. 
We introduced this prescription into Eq.~(\ref{CMLtau}), which becomes
 \begin{equation}
   x_{t+1}^i = (1-\epsilon) f(x_t^i) +  \epsilon 
   \sum_{r=1}^{N'} A(r) \Bigl( f(\hat{x}_{t}^{i-r}) +
     f(\hat{x}_{t}^{i+r})  \Bigr). 
   \label{CMLdelta}
 \end{equation}

\begin{center}
\begin{figure}[b!]
\includegraphics[width=0.9\linewidth,angle=0]{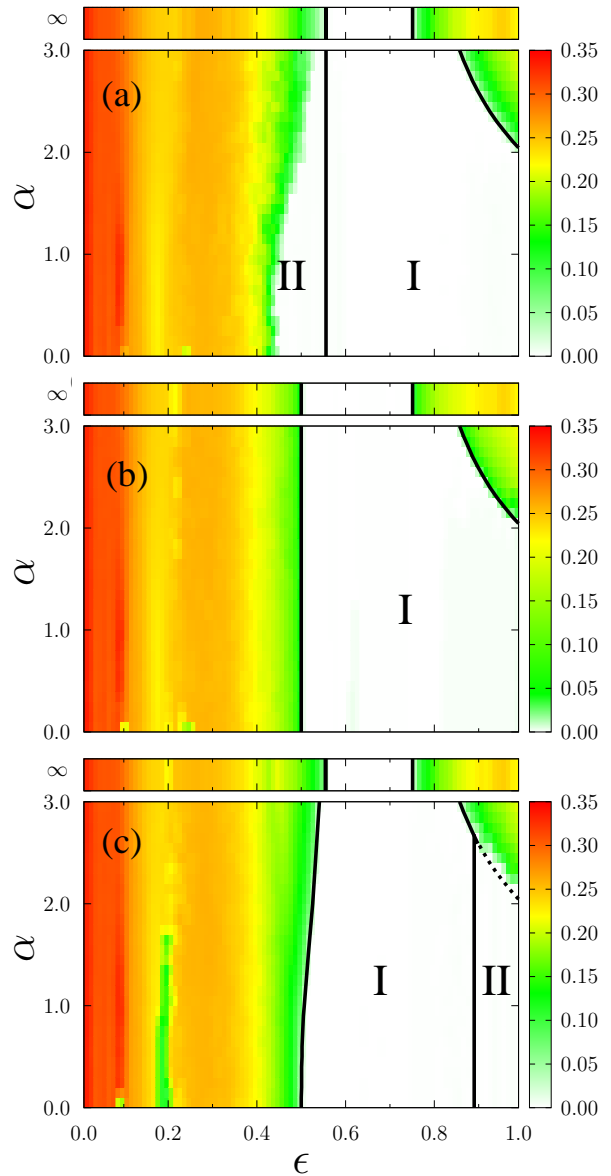}
\caption{(Color online)  {\bf Time-delayed dynamics}. 
	Phase diagrams, 	in  the parameter plane $\epsilon-\alpha$, showing 
	$\langle \sigma \rangle$  in color scale for different values of $\beta$ of the pseudo continuous dynamics 
($\hat{x}_t=\beta x_{t-1}+(1-\beta)x_{t-0}$): $\beta=0.45$ (a),	 $\beta=0.5$ (b) and $\beta=0.55$ (c). 
White regions indicate CS states, where $\langle \sigma \rangle=0$. 
The thick lines were obtained analytically through the stability  condition 
	$|\lambda|\le 1$, where $\lambda$ are the eigenvalues of $\mathbf{\tilde{F}}$ defined by Eq.~(\ref{tildeF}), 
as explained in the text.  
They delimit the region where the CS state is a fixed point ($x^\star=3/4$, subdomain  ``I''). 
While the subdomain   ``II'' corresponds to regular CS states of period larger than 1. 
	}
\label{fig:PDb}
\end{figure}
\end{center}

We observed that when $\beta$ is uniformly distributed in [0,1], the critical frontier is 
closer to  the asynchronous one than in the case of binary delays (not shown), but the similarity is enhanced when 
the distribution concentrates around the middle point and the  coincidence is  
perfect when $\beta=1/2$, as depicted in   Fig.~\ref{fig:PDb}, 
where we consider different values of $\beta \simeq 0.5$.  
We remark that outside the CS domain,  differential features emerge, 
as can be seen by comparing the color maps in Figs.~\ref{fig:PDa}.a  and ~\ref{fig:PDb}.b, 
with more complex structures in the delayed system. 
That is, the equivalence holds only for the CS domain when $\beta=0.5$.

The CS frontier for the system described by Eq.~(\ref{CMLdelta}) can be analytically determined as follows. 
First we cast the CML (\ref{CMLdelta})  in  Markovian form by the usual procedure of extending 
the phase space, such that $\mathbf{\delta \tilde{x}}_t=(\mathbf{\delta x}_t,\mathbf{\delta x}_{t-1})$. 
As a consequence, the matrix $\mathbf{\tilde{F}}$ which rules the linearized evolution around the fixed point, 
$\mathbf{\delta \tilde{x}}_t=\mathbf{\tilde{F}}\mathbf{\delta \tilde{x}}_{t-1}$, is the block matrix
\begin{equation}
 \mathbf{ \tilde{F}}  = \left( 
\begin{array}{c|c}
 [(1-\epsilon)\mathbb{1}+(1-\beta)\epsilon \mathbf{A}]f'    & \beta\epsilon \mathbf{A}f'  \\
\hline
                \mathbb{1}                          &\mathbb{0}  
\end{array} 
\right) ,
\label{tildeF}
\end{equation}
where  $f'\equiv f'(x*)=f'(3/4)=-2$. 
By setting $ \mathbf{ \tilde{F}}  ( u, v)^t = \lambda ( u, v)^t $, 
and substituting Eq.~(\ref{tildeF}), 
it can be easily shown that the eigenvalues $\lambda$ of the block matrix are related 
to the eigenvalues of $\mathbf{A}$, through the characteristic equation
\begin{equation}
\lambda^2 +2\lambda[  1-\epsilon +\epsilon(1-\beta)a_k] +2\beta\epsilon a_k =0.
\end{equation}
(Notice that, since this is a second order polynomial, two values of $\lambda$ arise for each value $a_k$.)
Of course,  if $\beta=0$, then the eigenvalues of $\mathbf{F}$ are recovered.
The condition $|\lambda|\le 1$ for all the eigenvalues furnishes the region of stability of the fixed point. 
For $\beta=1/2$, the frontier of the asynchronous case is exactly recovered, as can be seen in Fig.~\ref{fig:PDb}b.
The eigenvalues  $\lambda$ associated to the largest eigenvalue of $\mathbf{A}$, $a_N=1$, 
provide  the longitudinal stability,  
while the remaining ones furnish the transverse stability. 

In Fig.~\ref{fig:PDb}a-c, the white region corresponds to CS states ($\langle\sigma\rangle=0$). 
Inside that region,  the CS state is a fixed point ($x^\star=3/4$) in subdomain  ``I'', while the subdomain  
 ``II'' corresponds to regular CS states of period larger than 1.
Frontiers of the region where CS occurs at the fixed point are given by transverse stability condition, 
except those borders separating  regions I and II, in panels (Fig.~\ref{fig:PDb}.a and ~\ref{fig:PDb}.c), 
which are given by the longitudinal stability condition,  $1/4 \le \beta \epsilon \le 1/2$. 
If $\beta=0$, the fixed point cannot be stable for any coupling strength $\epsilon$, but above $\beta = 1/2$, 
there emerges an interval of 
values of $\epsilon$ for which the fixed point becomes stable.  
This explains the emergence of the stability of the locally unstable fixed point. \\

\section{Final remarks}
  
We found that binary discrete delays manage to mimic CS observed in asynchronous dynamics, 
but this occurs only if the interactions are sufficiently non-local. In contrast, 
pseudo-continuous delays reproduce exactly the asynchronous CS states, 
independently of the range of the interactions. 
This equivalence allows to embrace, in a unified frame, delayed and 
asynchronous dynamics, usually treated separately. 
The asynchronous case is usually analyzed heuristically, due to the mathematical difficulties involved, 
while there are analytical procedures to treat the delayed dynamics (by extending the phase space 
to turn the system Markovian). 
In this context, our finding provides an alternative way to determine the CS domains 
for an asynchronous update through a delayed one. 

As  side results, the outcomes put into evidence the interplay between 
the spatial and temporal dimensions in pattern formation,  and also have 
implications in control of chaos, as far as we found a time delayed scheme 
more efficient than discrete delays in the stabilization of  a locally unstable state, 
which can be understood along the lines discussed at the end of Sec.~\ref{sec:delayed}.

Despite we used a particular CML in  numerical examples, 
the analytical considerations apply to a general class of CMLs, 
following the form given by Eq.~(\ref{eq:CML}), with generic $A(r)$, and a chaotic unimodal map as local dynamics. 
However, the results might apply to a larger class of CMLs, e.g., including advective~\cite{lind04b} coupling or 
bistable local dynamics~\cite{lind04c}.

 {\bf Acknowledgments:} We acknowledge Brazilian agencies CNPq and FAPERJ for financial support.


\begin{thebibliography}{99}
 
\bibitem{PykovskyBook}
A. Pikovsky, M. Rosenblum,  J. Kurths,
  {\it Synchronization: a universal concept in nonlinear   sciences},  
	The Cambridge nonlinear science series 
  (Cambridge University Press, Cambridge,2001).

\bibitem{ManrubiaBook}
S.C.   Manrubia, A.S. Mikhailov, and D. Zanette, 
{\it Emergence of Dynamical Order. Synchronization
  Phenomena in Complex Systems} 
	(World Scientific, Singapore, 2004).

\bibitem{KuramotoBook}
Y.  Kuramoto, {\it Chemical Oscillations, Waves, and Turbulence} 
  (Springer, Berlin, 1984).

\bibitem{KanekoBook}
J. Crutchfield,  K. Kaneko,
{\it Theory and Applications of Coupled Map Lattices} 
  (World Scientific, Singapore, 1987).
	
\bibitem{HertzBook}
J. Hertz, A.   Krogh,  R.~G. Palmer, 
{\it Introduction to the Theory of   Neural Computation} 
(Addison-Wesley, Redwood, 1991).

\bibitem{Sinha2007}
M.D. Shrimali, S. Sinha,  K. Aihara,
  Phys. Rev. E {\bf 76}, {046212}   ({2007}).

\bibitem{Sinha_2000}
M. Mehta,  S. Sinha,
Chaos {\bf 10}, 350 (2000).

\bibitem{Lumer1994}
E. D. Lumer,  G. Nicolis,
Physica D {\bf 71},  440 (1994).
	
	\bibitem{nuno1} N. Crokidakis, V. H. Blanco,  C. Anteneodo, 
Phys. Rev. E {\bf 89}, 013310 (2014).
	
	\bibitem{nuno2}
	N. Crokidakis and C. Anteneodo, Phys. Rev. E {\bf 86}, 061127 (2012).
	
	
\bibitem{unstableFP1}
H. Atmanspacher,  T. Filk,  H. Scheingraber,
Eur. Phys. J. B {\bf 44},  229 (2005).

\bibitem{squeeze}
H. Atmanspacher,    H. Scheingraber,
I. J. Bifurcation and Chaos {\bf 15}, 1665 (2005).
 


\bibitem{Marti_PRL05}
C. Masoller, A.C. Mart\'{\i}, 
Phys. Rev. Lett. {\bf 94}, 134102 (2005).

\bibitem{Marti_EPJB05}
M. Ponce C., C. Masoller,  A. C. Mart\'{\i}, 
Eur.   Phys. J. B {\bf 67}, {83}  (2009).

 \bibitem{Marti_PRE05}
 A. C. Mart\'{\i},  M. Ponce, C. Masoller,  
Phys. Rev. E {\bf 72}, 066217 (2005).


\bibitem{delayed}
{\em Theme Issue 'Delayed complex systems'}, 
Phil. Trans. Royal Soc. A, 
compiled and edited by W. Just, A. Pelster, M. Schanz and E. Sch\"{o}ll (2009).


\bibitem{chemical1}
Y. Kuramoto, H. Nakao, Physica D {\bf 103}, 294 (1997).
 

\bibitem{Anteneodo2004}
C.Anteneodo,  A. Batista, R. Viana, 
Phys. Lett. A {\bf 326}, 227 (2004). 
	
	
	\bibitem{Anteneodo2006}
C.Anteneodo,  A. Batista, R. Viana, 
Physica D {\bf 223}, 270 (2006). 

\bibitem{gallas2004}
P.~G. Lind, J. Corte-Real, J.~A.~C. Gallas,  Phys. Rev. E {\bf 69}, 026209 (2004).

\bibitem{celia1998}
C. Anteneodo, C. Tsallis, 
Phys. Rev. Lett. {\bf 80},  {5313} (1998).

\bibitem{celia2000}
F. Tamarit, C. Anteneodo
Phys. Rev. Lett. {\bf 84},  {208} (2000).

 

\bibitem{lind04} P.G. Lind, J.A.C. Gallas, H.J. Herrmann, Phys. Rev. E {\bf 70}, 056207 (2004).
 

\bibitem{eckman-ruelle}
J.~P. Eckmann, D. Ruelle, 
Rev. Mod. Phys. {\bf 57} 617 (1985).

\bibitem{Benettin_01}
G. Benettin, L. Galgani, A. Giorgilli, and J.-M. Strelcyn,
Meccanica {\bf 15}, {9} (1980); {\em ibid.}  21  (1980).


 	
	
\bibitem{lind04b}
 P.G. Lind, J. Corte-Real, J.A.C. Gallas,  Phys. Rev. E {\bf 69}, 066206 (2004).

\bibitem{lind04c}
 P.G. Lind, S. Titz, T. Kuhlbrodt, J.A.M. Corte-Real, J. Kurths, J.A.C. Gallas, U. Feudel, 
Int. J. Bif. and Chaos {\bf 14}, 999 (2004).


\end{thebibliography}

\end{document}